\documentclass[aps,prl,twocolumn,superscriptaddress,showpacs]{revtex4}

\usepackage{graphicx}
\usepackage{dcolumn}
\usepackage{bm}

\begin{document}
\newlength{\ogltemp}

\def\met{\settowidth{\ogltemp}{\big\slash}%
\hspace{0.5mm}\big\slash\hspace{-1.4\ogltemp}{E_T}%
}
\def\vmet{\settowidth{\ogltemp}{\big\slash}%
\hspace{0.5mm}\big\slash\hspace{-1.4\ogltemp}{\vec{E}_T}%
}

\def\mel{\big\slash\hspace{-1.0ex}{\ell}}

\def\gev{\,\,\mathrm{GeV}\,}
\def\tev{\,\,\mathrm{TeV}\,}
\def\mev{\,\,\mathrm{MeV}\,}
\def\gevc{\,\,\mathrm{GeV}/c\,}
\def\gevm{\,\,\mathrm{GeV}/c^2\,}
\newcommand{\err}[2]{$^{#1}_{#2}$}
\newcommand{\serr}[2]{$^{+#1}_{-#2}$}
\newcommand{\boldit}[1]{\textbf{\emph{#1}}}

\preprint{Draft 6}

\title{Search for the Higgs boson in events with missing transverse energy and \boldit{b} quark jets produced in \boldmath{$p\bar{p}$} collisions at \boldmath{$\sqrt{s}=1.96$}~TeV}

%


\affiliation{Institute of Physics, Academia Sinica, Taipei, Taiwan 11529, Republic of China} 
\affiliation{Argonne National Laboratory, Argonne, Illinois 60439} 
\affiliation{Institut de Fisica d'Altes Energies, Universitat Autonoma de Barcelona, E-08193, Bellaterra (Barcelona), Spain} 
\affiliation{Baylor University, Waco, Texas  76798} 
\affiliation{Istituto Nazionale di Fisica Nucleare, University of Bologna, I-40127 Bologna, Italy} 
\affiliation{Brandeis University, Waltham, Massachusetts 02254} 
\affiliation{University of California, Davis, Davis, California  95616} 
\affiliation{University of California, Los Angeles, Los Angeles, California  90024} 
\affiliation{University of California, San Diego, La Jolla, California  92093} 
\affiliation{University of California, Santa Barbara, Santa Barbara, California 93106} 
\affiliation{Instituto de Fisica de Cantabria, CSIC-University of Cantabria, 39005 Santander, Spain} 
\affiliation{Carnegie Mellon University, Pittsburgh, PA  15213} 
\affiliation{Enrico Fermi Institute, University of Chicago, Chicago, Illinois 60637} 
\affiliation{Comenius University, 842 48 Bratislava, Slovakia; Institute of Experimental Physics, 040 01 Kosice, Slovakia} 
\affiliation{Joint Institute for Nuclear Research, RU-141980 Dubna, Russia} 
\affiliation{Duke University, Durham, North Carolina  27708} 
\affiliation{Fermi National Accelerator Laboratory, Batavia, Illinois 60510} 
\affiliation{University of Florida, Gainesville, Florida  32611} 
\affiliation{Laboratori Nazionali di Frascati, Istituto Nazionale di Fisica Nucleare, I-00044 Frascati, Italy} 
\affiliation{University of Geneva, CH-1211 Geneva 4, Switzerland} 
\affiliation{Glasgow University, Glasgow G12 8QQ, United Kingdom} 
\affiliation{Harvard University, Cambridge, Massachusetts 02138} 
\affiliation{Division of High Energy Physics, Department of Physics, University of Helsinki and Helsinki Institute of Physics, FIN-00014, Helsinki, Finland} 
\affiliation{University of Illinois, Urbana, Illinois 61801} 
\affiliation{The Johns Hopkins University, Baltimore, Maryland 21218} 
\affiliation{Institut f\"{u}r Experimentelle Kernphysik, Universit\"{a}t Karlsruhe, 76128 Karlsruhe, Germany} 
\affiliation{Center for High Energy Physics: Kyungpook National University, Daegu 702-701, Korea; Seoul National University, Seoul 151-742, Korea; Sungkyunkwan University, Suwon 440-746, Korea; Korea Institute of Science and Technology Information, Daejeon, 305-806, Korea; Chonnam National University, Gwangju, 500-757, Korea} 
\affiliation{Ernest Orlando Lawrence Berkeley National Laboratory, Berkeley, California 94720} 
\affiliation{University of Liverpool, Liverpool L69 7ZE, United Kingdom} 
\affiliation{University College London, London WC1E 6BT, United Kingdom} 
\affiliation{Centro de Investigaciones Energeticas Medioambientales y Tecnologicas, E-28040 Madrid, Spain} 
\affiliation{Massachusetts Institute of Technology, Cambridge, Massachusetts  02139} 
\affiliation{Institute of Particle Physics: McGill University, Montr\'{e}al, Canada H3A~2T8; and University of Toronto, Toronto, Canada M5S~1A7} 
\affiliation{University of Michigan, Ann Arbor, Michigan 48109} 
\affiliation{Michigan State University, East Lansing, Michigan  48824} 
\affiliation{University of New Mexico, Albuquerque, New Mexico 87131} 
\affiliation{Northwestern University, Evanston, Illinois  60208} 
\affiliation{The Ohio State University, Columbus, Ohio  43210} 
\affiliation{Okayama University, Okayama 700-8530, Japan} 
\affiliation{Osaka City University, Osaka 588, Japan} 
\affiliation{University of Oxford, Oxford OX1 3RH, United Kingdom} 
\affiliation{University of Padova, Istituto Nazionale di Fisica Nucleare, Sezione di Padova-Trento, I-35131 Padova, Italy} 
\affiliation{LPNHE, Universite Pierre et Marie Curie/IN2P3-CNRS, UMR7585, Paris, F-75252 France} 
\affiliation{University of Pennsylvania, Philadelphia, Pennsylvania 19104} 
\affiliation{Istituto Nazionale di Fisica Nucleare Pisa, Universities of Pisa, Siena and Scuola Normale Superiore, I-56127 Pisa, Italy} 
\affiliation{University of Pittsburgh, Pittsburgh, Pennsylvania 15260} 
\affiliation{Purdue University, West Lafayette, Indiana 47907} 
\affiliation{University of Rochester, Rochester, New York 14627} 
\affiliation{The Rockefeller University, New York, New York 10021} 
\affiliation{Istituto Nazionale di Fisica Nucleare, Sezione di Roma 1, University of Rome ``La Sapienza," I-00185 Roma, Italy} 
\affiliation{Rutgers University, Piscataway, New Jersey 08855} 
\affiliation{Texas A\&M University, College Station, Texas 77843} 
\affiliation{Istituto Nazionale di Fisica Nucleare, University of Trieste/\ Udine, Italy} 
\affiliation{University of Tsukuba, Tsukuba, Ibaraki 305, Japan} 
\affiliation{Tufts University, Medford, Massachusetts 02155} 
\affiliation{Waseda University, Tokyo 169, Japan} 
\affiliation{Wayne State University, Detroit, Michigan  48201} 
\affiliation{University of Wisconsin, Madison, Wisconsin 53706} 
\affiliation{Yale University, New Haven, Connecticut 06520} 
\author{T.~Aaltonen}
\affiliation{Division of High Energy Physics, Department of Physics, University of Helsinki and Helsinki Institute of Physics, FIN-00014, Helsinki, Finland}
\author{J.~Adelman}
\affiliation{Enrico Fermi Institute, University of Chicago, Chicago, Illinois 60637}
\author{T.~Akimoto}
\affiliation{University of Tsukuba, Tsukuba, Ibaraki 305, Japan}
\author{M.G.~Albrow}
\affiliation{Fermi National Accelerator Laboratory, Batavia, Illinois 60510}
\author{B.~\'{A}lvarez~Gonz\'{a}lez}
\affiliation{Instituto de Fisica de Cantabria, CSIC-University of Cantabria, 39005 Santander, Spain}
\author{S.~Amerio}
\affiliation{University of Padova, Istituto Nazionale di Fisica Nucleare, Sezione di Padova-Trento, I-35131 Padova, Italy}
\author{D.~Amidei}
\affiliation{University of Michigan, Ann Arbor, Michigan 48109}
\author{A.~Anastassov}
\affiliation{Rutgers University, Piscataway, New Jersey 08855}
\author{A.~Annovi}
\affiliation{Laboratori Nazionali di Frascati, Istituto Nazionale di Fisica Nucleare, I-00044 Frascati, Italy}
\author{J.~Antos}
\affiliation{Comenius University, 842 48 Bratislava, Slovakia; Institute of Experimental Physics, 040 01 Kosice, Slovakia}
% ...
\collaboration{CDF Collaboration\footnote{With visitors from $^a$University of Athens, 15784 Athens, Greece, 
$^b$Chinese Academy of Sciences, Beijing 100864, China, 
$^c$University of Bristol, Bristol BS8 1TL, United Kingdom, 
$^d$University Libre de Bruxelles, B-1050 Brussels, Belgium, 
$^e$University of California Irvine, Irvine, CA  92697, 
$^f$University of California Santa Cruz, Santa Cruz, CA  95064, 
$^g$Cornell University, Ithaca, NY  14853, 
$^h$University of Cyprus, Nicosia CY-1678, Cyprus, 
$^i$University College Dublin, Dublin 4, Ireland, 
$^j$University of Edinburgh, Edinburgh EH9 3JZ, United Kingdom, 
$^k$University of Heidelberg, D-69120 Heidelberg, Germany, 
$^l$Universidad Iberoamericana, Mexico D.F., Mexico, 
$^m$University of Manchester, Manchester M13 9PL, England, 
$^n$Nagasaki Institute of Applied Science, Nagasaki, Japan, 
$^o$University de Oviedo, E-33007 Oviedo, Spain, 
$^p$Queen Mary, University of London, London, E1 4NS, England, 
$^q$Texas Tech University, Lubbock, TX  79409, 
$^r$IFIC(CSIC-Universitat de Valencia), 46071 Valencia, Spain, 
}}
\noaffiliation


\date{\today}



\begin{abstract}

We search for the standard model Higgs boson produced in association with an electroweak 
vector boson in events with no
identified charged leptons, large imbalance in transverse momentum, and two jets where at least
one contains a secondary vertex consistent with the decay of $b$ hadrons.
We use $\sim$1~fb$^{-1}$ integrated luminosity of
{\bf $p\bar{p}$} collisions at {\bf $\sqrt{s}=1.96$}~TeV recorded
by the CDF II experiment at the Tevatron.   We find 268 (16) single
(double) $b$-tagged candidate events, where $248 \pm 43$ ($14.4\pm 2.7$)
are expected from standard model background processes. We place $95\%$
confidence level upper limits on the Higgs boson production cross
section for several Higgs boson masses ranging from 110~GeV/$c^2$ to 140~GeV/$c^2$.
For a mass of 115~GeV/$c^{2}$ the observed (expected) limit is 20.4 (14.2) times the standard model prediction.

\end{abstract}

\pacs{12.60.Jv}
\maketitle


The Higgs boson is an essential element of the standard model (SM) of
particles and their interactions explaining the mass of elementary particles
and playing a key role in the mechanism of electroweak symmetry
breaking. Precision measurements of the $W$ boson and top quark masses
suggest that the mass of the SM Higgs boson
should be less than 182~GeV/c$^2$ (95\%\,C.L.)\,\cite{precisemeas} and therefore 
kinematically within reach at the Tevatron. Previous
searches for the SM Higgs boson set limits on its production
cross section at $\sqrt{s}=1.8$~TeV and $\sqrt{s}=1.96$~TeV
\protect\cite{prevhiggs}.
The CDF Collaboration is now performing searches for the low mass
Higgs boson produced in association with an electroweak vector
boson $V$ ($Z$ or $W$) followed by the decay $H\rightarrow b\bar{b}$ in
several final states which are optimized separately and then
combined. In this Letter, we describe the first search for the SM
Higgs boson with a trigger based on an imbalance in energy in the
plane transverse to the beam sensitive to both $ZH$ and $WH$
associated productions. In order to avoid overlap with the search
for $WH$ in a final state containing a high-energy electron or muon
candidate, large missing transverse energy, and one or two $b$
jets, we veto events with high energy isolated leptons. Therefore,
we search for $VH$ events where $H\rightarrow b\bar{b}$ and
either $Z\rightarrow \nu \bar{\nu}$ or $W\rightarrow \mel\nu$, where
$\mel$ indicates that the lepton was not identified in
the detector.   The signatures of these decay modes are final
states with two $b$ jets from the hadronization of the $b$ quarks,
missing transverse energy,
and no isolated leptons. In this search we use an integrated luminosity
of $0.97 \pm0.06$~fb$^{-1}$ collected in Run II of
the Tevatron. After describing the experiment, we introduce our
event selection, major backgrounds, and systematic uncertainties,
and present the results.

CDF II is a multipurpose detector described in detail
elsewhere~\protect\cite{CDFII}. Here, we briefly describe the
detector components used in this analysis. We use a cylindrical coordinate
system around the proton beam axis in which $\theta$ and $\phi$ are the
polar and azimuthal angles, respectively, the
pseudorapidity $\eta$ is defined as $\eta
=-\ln(\tan(\frac{\theta}{2}))$. The transverse momentum and energy
of a particle are defined as $p_T=p \sin\theta$ and \mbox{$E_T=E \sin\theta$},
respectively. The tracking system consists of a cylindrical
open-cell drift chamber and silicon microstrip detectors in a
1.4~T magnetic field parallel to the beam axis. The silicon
detectors~\protect\cite{SVXII} provide tracking information for $|\eta|<2$ and
are used to detect collision and decay points. The
drift chamber~\protect\cite{COT} surrounds the silicon detectors
and covers the central rapidity region $|\eta|<1$.

The energies of
electrons and jets are measured in calorimeters which cover the region 
$|\eta|<3.6$ and are segmented
into towers pointing toward the center of the detector.
Jets are reconstructed from energy depositions in the
calorimeter towers using a jet clustering cone
algorithm~\protect\cite{jes} with a cone size of
radius \mbox{$\Delta R = \sqrt{(\Delta\phi)^2+(\Delta \eta)^2} = 0.4$}.
Corrections are applied to account for effects that can
cause mismeasurements in the jet energy such as non-linear calorimeter
response, multiple beam interactions, or displacement of the event vertex
from the nominal  $z=0$ position~\protect\cite{jes}. 
The $\met$ is defined as the magnitude of $\vmet = -\sum_{i}
E^i\hat{n}_i$, where $\hat{n}_i$ is a unit vector parallel to the 
transverse component of the vector pointing at the $i^{th}$ calorimeter tower,
and $E^i$ is the energy therein. Both magnitude and
direction of the $\vmet$ are recomputed after the jet energies are corrected. 

We use loose
electron and muon identification criteria in order to veto
events with at least one high $p_T$ isolated lepton. Candidate muons are
identified with $p_T>10\gevc$ isolated tracks which leave energy
in the calorimeters that is consistent with a minimum ionizing
particle. A track is called isolated if the total calorimeter
energy in a 0.4 radius cone around the track is less than 10\% of
the track $p_T$. Candidate electrons are clustered energy deposits
in the electromagnetic calorimeter with $E_T>10$~GeV which have
electromagnetic to hadronic energy ratio and shower shape
compatible with electrons and are associated with an isolated track.

The events used in this search are selected by a three-level
trigger system. The Level 1 trigger requires $\met \ge 25\gev$,
where the $\met$ is determined using calorimeter trigger towers
with $E_T > 1$~GeV. The Level 2 trigger selects events with two
jet clusters of $E_T>10$~GeV, where one of the clusters is
required to be in the region $|\eta|<1.1$. The Level $3$
requirement is $\met > 35$~GeV.  After selections to remove
accelerator-produced and detector-related background, as well as
cosmic ray events, 4.3 million events remain. This is about 60\%
of the initial sample. Further selection criteria are defined
using corrected observables in order to search in a region where
the trigger is fully efficient, but the sensitivity of the search
is not compromised.  We require the events to have $\met
> 50$~GeV and to contain exactly two jets with $E_T > 20$~GeV and
 $|\eta| < 2.4$ where the azimuthal separation of the
two jets is $\Delta\phi(\vec{E}_{1,T}, \vec{E}_{2,T})>$~1.0 rad.
Moreover one of the two jets must be central with $|\eta| < 0.9$,
and the jet of highest (second highest) energy in the event must
have $E_{T,1}>35$~GeV ($E_{T,2}>25$~GeV). The trigger efficiency
for such events is $0.97 \pm 0.03 $ \protect\cite{viktor}.

We require at least one jet to be identified as a $b$ jet by the
$b$-tagging algorithm {\sc secvtx}~\protect\cite{btag}. This
algorithm looks for tracks within jets that form a secondary
vertex significantly displaced
from the primary interaction point. 
The tagging efficiency for $b$ jets of 50~GeV is $\sim$42\%,
the misidentification rate for 50~GeV  light quark ($u$, $d$, $s$) and gluon jets is $\sim$1\%
\protect\cite{btag}.

In order to avoid potential bias in the search, we test our understanding of
the background in control regions that are defined {\it a
priori}. The observed events are divided into three non-overlapping parts, two control and one
signal regions, which have different
event topology and isolated high-$p_T$ lepton multiplicity. The backgrounds in the
three regions have
contributions from multi-jet, $t\bar t$, $W$ plus jets, $Z$ plus jets, and electroweak
($WW$, $WZ$, and $ZZ$) processes.
Control region 1 (CR1) is defined such that it is dominated by the multi-jet background. The
events in this region do not contain any identified leptons, and the separation between the $\vmet$ and the secondary jet ($\Delta\phi(\vec{E}_{2,T}, \vmet)$) is required to be less than 0.4.
Control region 2 (CR2) is selected by
requiring at least one lepton and $\Delta\phi(\vec{E}_{2,T},\vmet)\geq$~0.4. This
region is sensitive to electroweak  and top quark decays.
The signal region contains events with
no identified high-$p_T$ leptons,
$\Delta\phi(\vec{E}_{1,T},\vmet)\geq$~0.4, and $\Delta\phi(\vec{E}_{2,T},\vmet)\geq$~0.4.

Agreement between the number of events expected and observed
due to different processes is presented in Table~\ref{tab:table1}
for single and double $b$-tagged events in CR1 and CR2. The signal
and the physics backgrounds have been evaluated using {\sc
pythia}~\protect\cite{pythia} Monte Carlo generation (MC) followed
by the simulation of the CDF II detector. QCD light jet production has a
cross section about nine orders of magnitude greater than the theoretical
expectation for the signal before the $b$-tag requirement. Although this process
generally does not have intrinsic $\met$, mismeasured jets can
cause an imbalance in the total $E_T$ and fake the signal if one
of the jets is misidentified as a $b$ jet (mistag). Furthermore,
$b$ quark pair production yields $b$ jets and  $\met$ if one $b$ quark
undergoes a semi-leptonic decay. In both cases, the $\vmet$ tends
to be aligned parallel or anti-parallel to the first or second
most energetic jet. We use MC simulation only to evaluate QCD processes which
yield events with a $b$ or $c$ quark (heavy quark, h.q.) pair. This background is denoted as
QCD\ h.q.\ in Table~\ref{tab:table1}. The MC cross section
for heavy quark jet production is scaled to the difference
between the observed data and the predicted contribution of the
other backgrounds.  We determine the normalization factors,
$k_{\mathrm{QCD}}$, before the data are divided into control and signal
regions to be $1.30 \pm 0.04$ and $1.47 \pm 0.07$ for events
with one and two $b$-tagged jets, respectively \protect\cite{viktor}.
The background due to light quarks being misidentified as $b$ jets is denoted as mistags in Table~\ref{tab:table1} and is determined using a $b$-jet misidentification rate that is measured from inclusive jet samples~\protect\cite{btag}. 
The $t\bar{t}$ cross section of
$7.3 \pm 0.8$~pb is obtained from the combined CDF measurements \protect\cite{top}.
A small contribution by single top quark production is included, normalized with theoretical
NLO production cross sections~\protect\cite{singletops,singletopt}.
The electroweak backgrounds were generated at leading order with {\sc pythia}
and increased by a factor of 1.4 to account for higher order effects \protect\cite{wznorm, dbnorm}.


\renewcommand{\arraystretch}{0.9}
\begin{table*}[b,h,t]
\caption{Comparison of the total number of expected and observed single and double $b$-tagged events in the control regions and in the optimized signal region.
}
    \label{tab:table1}
    \begin{ruledtabular}
    \begin{tabular}{lcccccc}
& \multicolumn{2}{c}{Control Region 1} & \multicolumn{2}{c}{Control Region 2} & \multicolumn{2}{c}{Signal Region} \\ \hline
Lepton &\multicolumn{2}{c}{Vetoed} &\multicolumn{2}{c}{Identified} &\multicolumn{2}{c}{Vetoed}\\
& \multicolumn{2}{c}{$\Delta\phi(\vec{E}_{2,T}, \vmet)<$0.4 } & \multicolumn{2}{c}{$\Delta\phi(\vec{E}_{2,T}, \vmet)\geq$0.4} & \multicolumn{2}{c}{See text} \\
$B$-tag&  Single & Double  & Single & Double & Single & Double \\
\hline
QCD h.q. & $14868 \pm 1820$ & $1175 \pm 179$  & $61 \pm 25$     & $3.2 \pm 1.7$
& $93 \pm 23$        & $3.74 \pm 1.27$ \\
Top       & $4.0 \pm 0.6$    & $0.8 \pm 0.2$   & $98 \pm 14$     & $24.0 \pm 3.8$
& $27.3 \pm 3.8$     & $4.88 \pm 0.80$ \\
Di-boson  & $0.7 \pm 0.1$    & $0.04 \pm 0.02$ & $11.5 \pm 2.1$  & $1.0  \pm 0.2$
& $7.0 \pm 1.4$      & $0.79 \pm 0.19$  \\
$W$+ h.q.   & $14.0 \pm 6.7$   & $0.3 \pm 0.2$   & $63 \pm 28$     & $3.9 \pm 1.8$
& $33.4 \pm 16.2$    & $1.65 \pm 0.86$ \\
$Z$+ h.q.  & $5.3 \pm 2.2$    & $0.3 \pm 0.2$   & $12.9 \pm 5.9$  & $0.5 \pm 0.4$
& $18.3 \pm 8.1$ & $1.67 \pm 0.77$  \\
Mistags   & $3450 \pm 433$   & $73 \pm 14$     & $85\pm 11$      & $2.2  \pm 0.6$
& $69 \pm 9$        & $1.64 \pm 0.48$ \\  \hline
Expected  & $18342\pm2031$   & $1249 \pm 203$  & $331\pm 51$     & $34.8 \pm 5.9$
& $248 \pm 43$       & $14.4 \pm 2.7$\\
Observed  &  18588           &  1251           &  373            &  28
&  268               &  16         \\
    \end{tabular}
    \end{ruledtabular}
\end{table*}

\begin{figure*}[t,h]
\includegraphics*[width=7cm]{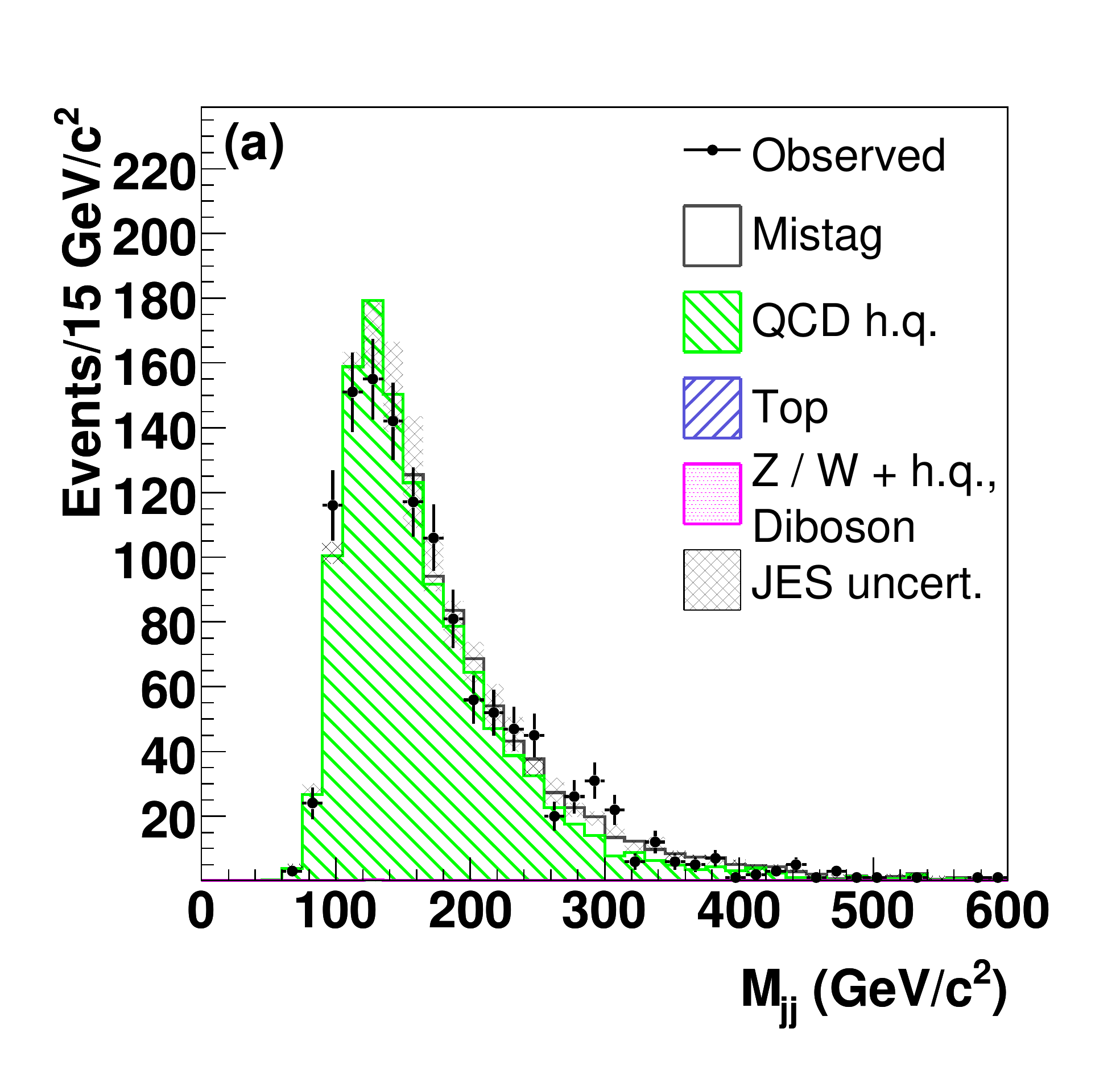}
\includegraphics*[width=7cm]{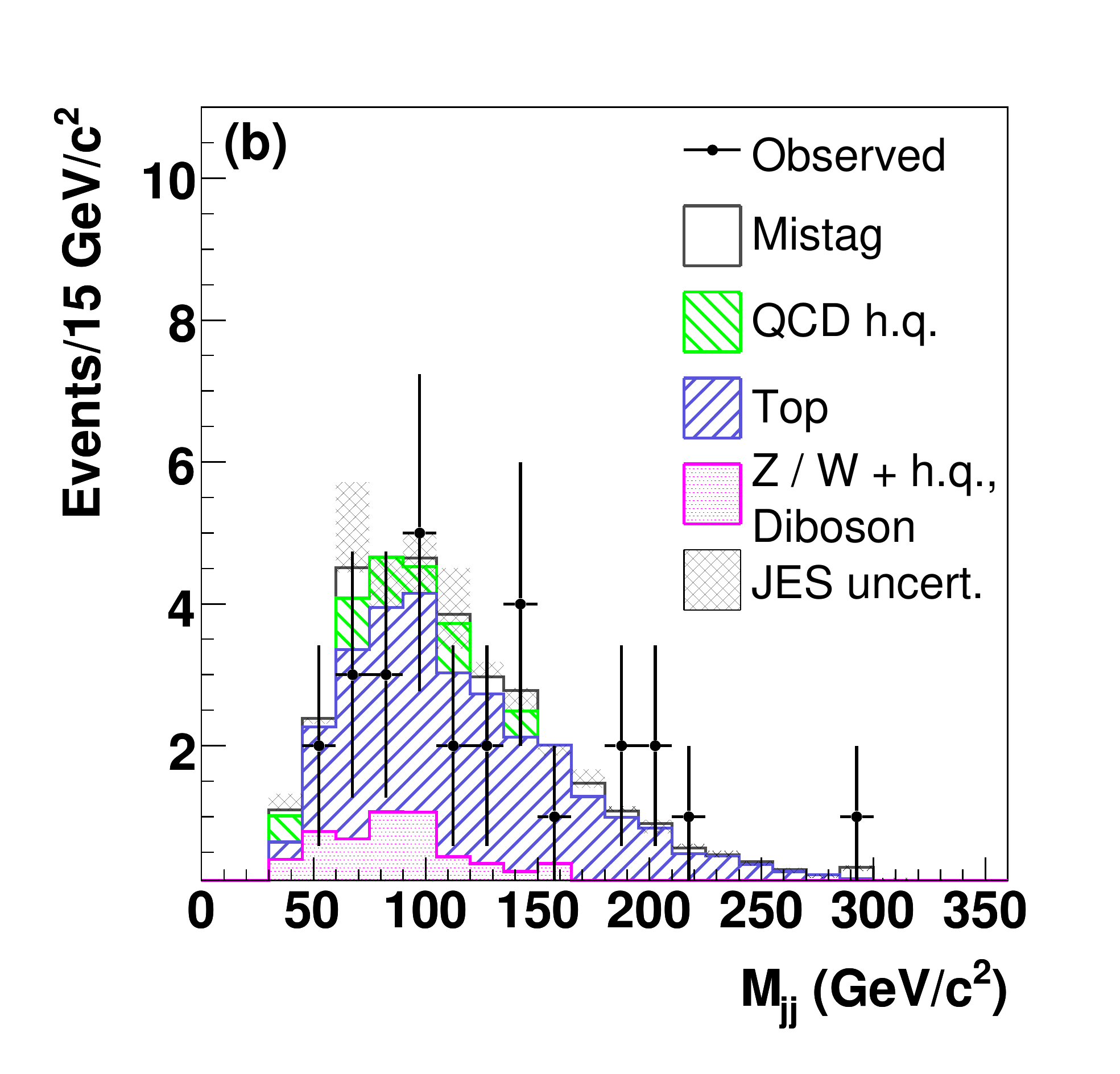}
\caption{Dijet invariant mass distribution for double-tagged events in (a) CR 1 and (b)
CR 2} \label{fig:spectrums}
\end{figure*}

Simulated distributions of kinematic variables such as the jet $E_T$, the
dijet invariant mass, and the $\met$ have been found to
be in agreement with observations in the control regions \protect\cite{viktor}. 
In Fig.\,\ref{fig:spectrums} (a) and \ref{fig:spectrums} (b) we
show the dijet invariant mass distributions when both jets are $b$-tagged
in control regions CR1 and CR2, respectively.

We optimize the sensitivity to Higgs boson production by maximizing the
ratio of $N_{\rm Higgs}/\sqrt{N_{\rm bck}}$ where $N_{\rm Higgs}$ and
$N_{\rm bck}$ are the number of predicted Higgs boson and background events
in the signal region.
The final selection requires $\Delta\phi(\vec{E}_{1,T},\vmet) >0.8$,
$\big\slash\hspace{-1.6ex}{H_T}/{H_T}>0.45$, $E_{T,1}>60
\gev$, and $\met>70 \gev$. The variable $H_T$ ($\big\slash\hspace{-1.6ex}{H_T}$)
is the scalar (the magnitude of the vectorial) sum of the two jet energies in the
transverse plane. The number of
background events after the final selection is shown in Table
\ref{tab:table1}. For a Higgs boson of mass 115~GeV/$c^{2}$ we expect a total
of 1.2 (0.4) events with one (two) $b$-tagged jets \protect\cite{Higgs}.




The total systematic uncertainties on the background predictions
are 17\% and 19\% in the single and double tag events, respectively
\protect\cite{viktor}. One of the dominant correlated
uncertainties for the background and signal predictions is due to
the jet energy scale (JES) \protect\cite{jes}. The JES uncertainty affects selection
efficiencies non-uniformly over kinematic distributions leading to
uncertainties in the distributions' shapes. The resulting systematic
uncertainties are typically between 10\% and 20\% for multi-jet
events depending on the kinematic selections, 15\% for dibosons,
26\% for $W$ plus h.q., 17\% for $Z$ plus h.q., 
1\% for top quark, and 8\% for signal events. The systematic
uncertainty on the $b$-tagging efficiency is 4.3\% for single and
8.6\% for double $b$-tagged events. There is a 3\% uncertainty on the
trigger efficiency, 2\% uncertainty on the signal and background
acceptances due to uncertainties in the parton distribution functions,
and 6\% on luminosity. The
dominant uncorrelated systematic uncertainties are due to the
cross sections and the statistics of the simulated samples. We
assign uncertainties of 11\% to the top, 11.5\% to the diboson,
and 40\% to the $W$ plus h.q. and $Z$ plus h.q. cross sections.
Statistical uncertainties due to Monte Carlo sample sizes are 20\%-32\%
on the multi-jet, 20\% on the $W$ plus h.q., and 11\% on the $Z$ plus h.q.
predictions.
Correlated and uncorrelated uncertainties are evaluated separately and
combined in quadrature.


\begin{figure*}[t,h]
        \includegraphics[width=7cm]{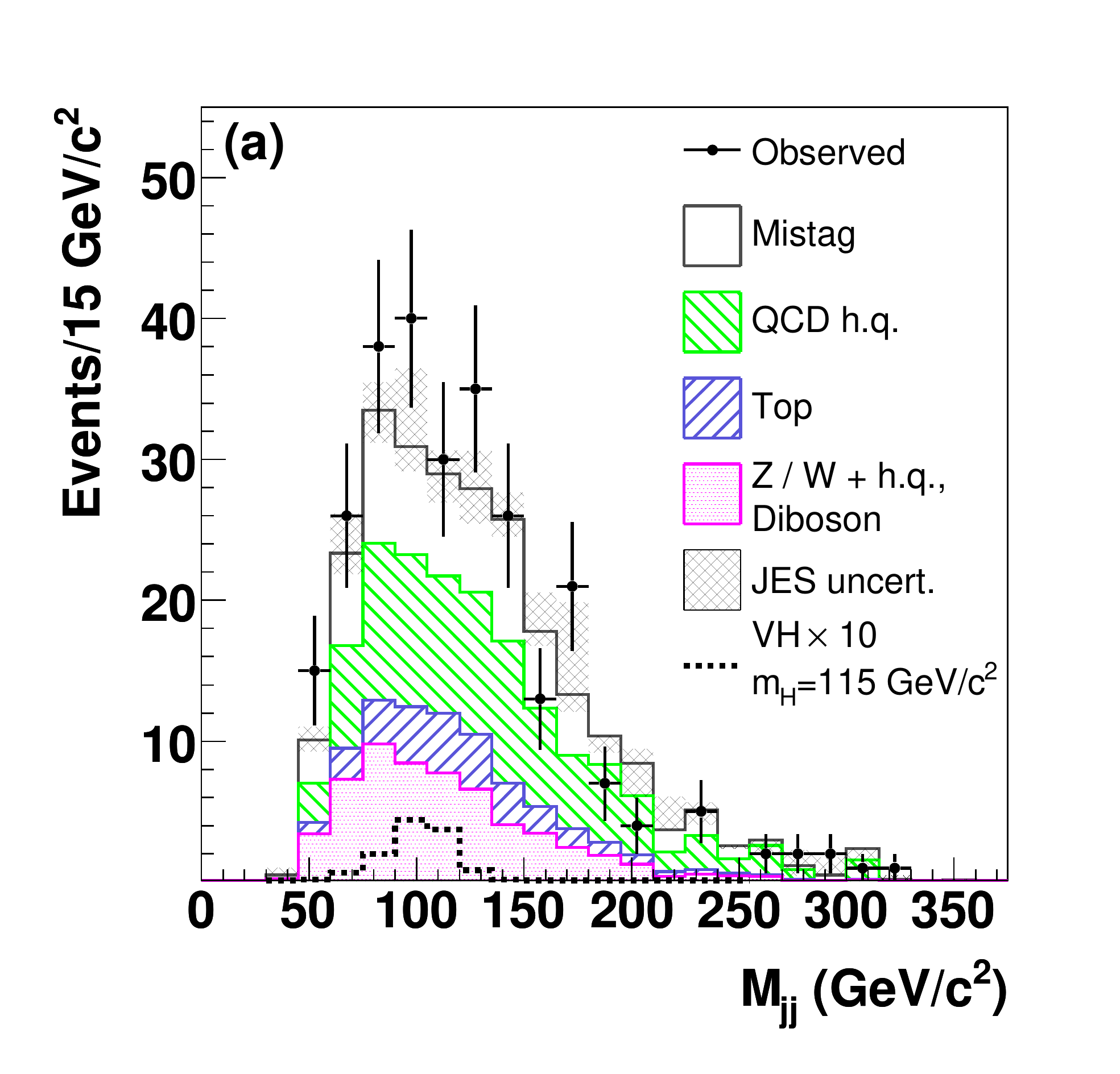}
        \includegraphics[width=7cm]{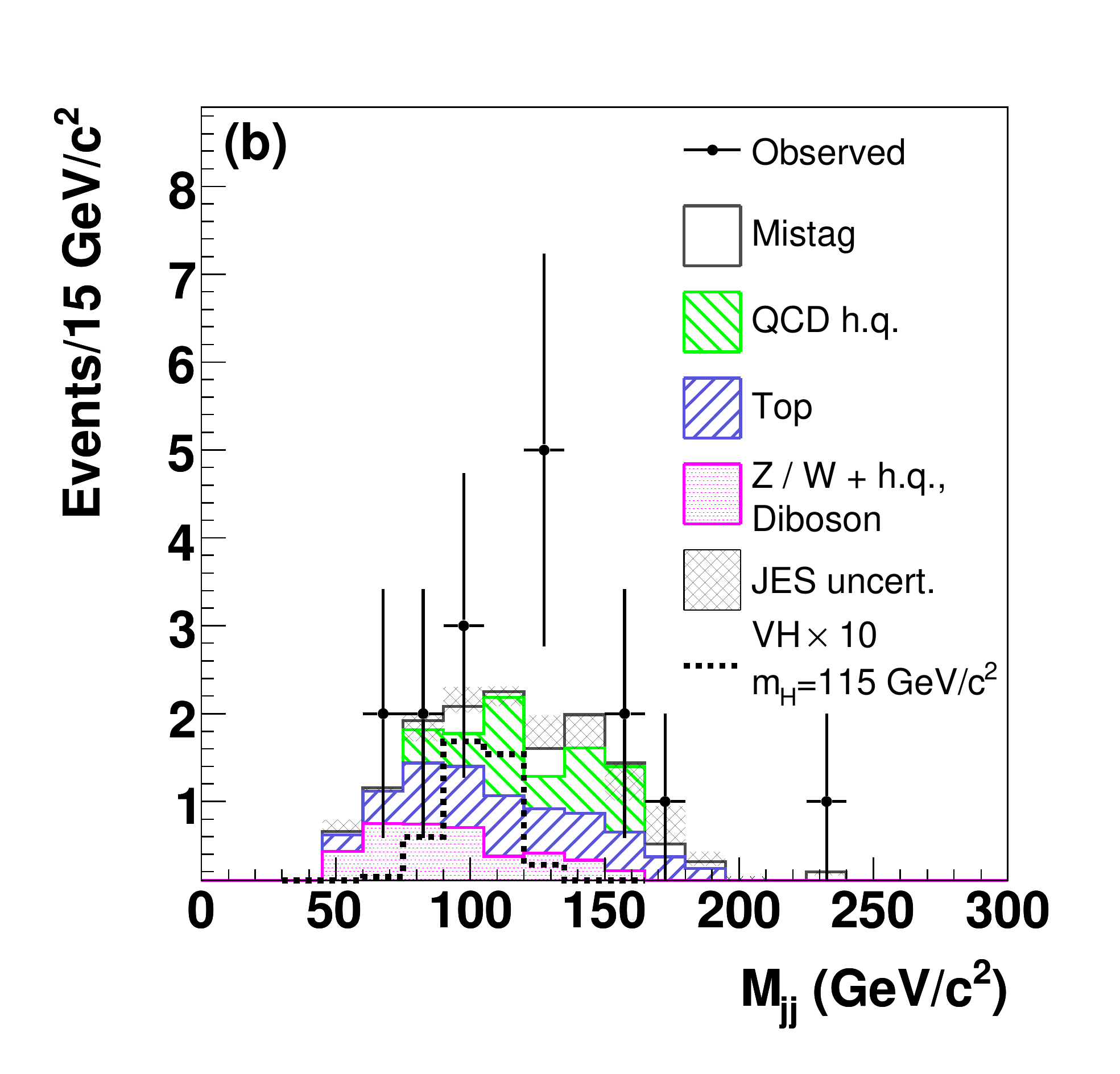}
        \caption{Dijet invariant mass in the signal region for (a) single-tagged and (b) double-tagged events.
                        The dotted line overlaid on the background shows 10 times the expected yield
                        of 115~GeV/$c^{2}$ Higgs bosons.}
    \label{fig:SIG_mjj_Final}
\end{figure*}

The signal region is analyzed after all the background predictions
and selection requirements were determined. The results are summarized in Table~\ref{tab:table1}. Requiring one (two)
$b$-tagged jets we observe 268 (16) events, where $248 \pm 43$ ($14.4\pm2.7$)
are expected. The dijet invariant mass distributions in the signal region when one and two jets are $b$-tagged are shown in Fig.~\ref{fig:SIG_mjj_Final}.

Since no significant excess is observed, we compute 95\% C.L. upper limits for
the Higgs boson production cross section times the branching fraction when the Higgs boson is produced in association with a $Z$ or $W$ boson and decays to two
$b$ quarks while the $Z$ and $W$ yield large $\met$ and no identified leptons. The limits are computed
using the Bayesian likelihood method~\protect\cite{limits} with
flat prior probability for the signal cross section and Gaussian
priors for the uncertainties on acceptance and backgrounds. The
likelihood quantifies the agreement between the dijet mass spectrum in the observed data
and the sum of the modeled background processes and $VH$ signal as a function of 
the Higgs boson mass. We combine
the channels, such as single $b$-tag, double $b$-tag, $ZH$, and $WH$ by taking
the product of their likelihoods and simultaneously varying the
correlated uncertainties. Table \ref{tab:table2} summarizes the
combined single and double $b$-tag upper limits separately for $ZH$ and
$WH$, and the combined $VH$ upper limits. The result is also expressed as the
ratio of the upper limit to the SM cross sections at NLO as a
function of the Higgs boson mass. The observed limits agree with the expected ones at the 
$1\,\sigma$ level.

\renewcommand{\arraystretch}{1.0}
\begin{table*}[t,h]
\caption{The expected cross section limits of the $ZH$ and
$WH$ processes, and the expected and observed combined ($VH$)
cross section limits when $H\rightarrow b\bar{b}$. The last column gives the ratio of the observed limit with respect to the SM cross section. }\label{tab:table2}
\begin{ruledtabular}
\begin{tabular}{lccccc}
Higgs mass &
$\sigma(ZH)\times {\rm Br}(H\rightarrow b\bar{b})$ &
$\sigma(WH)\times {\rm Br}(H\rightarrow b\bar{b})$ &
\multicolumn{2}{c}{$\sigma(VH)\times {\rm Br}(H\rightarrow b\bar{b})$}   & Ratio      \\
   & Expected  &    Expected  &Expected &Observed & Observed\\
(GeV/$c^{2}$) & (pb) & (pb)  &(pb)&(pb)& \\
\hline
110 & 2.1\serr{0.9}{0.7} & 4.8\serr{2.7}{1.2} & 3.1\serr{1.8}{0.9} & 4.7 & 17.9\\
115 & 2.1\serr{0.8}{0.7} & 4.6\serr{2.2}{1.4} & 3.0\serr{1.4}{0.9} & 4.4 & 20.4\\
120 & 1.8\serr{0.8}{0.6} & 4.4\serr{1.9}{1.4} & 2.9\serr{1.3}{0.9} & 4.3 & 24.8\\
125 & 1.7\serr{0.7}{0.6} & 3.9\serr{2.0}{1.0} & 2.5\serr{1.0}{0.6} & 4.0 & 30.2\\
130 & 1.6\serr{0.9}{0.5} & 3.4\serr{1.2}{1.0} & 2.4\serr{1.1}{0.8} & 3.8 & 38.3 \\
135 & 1.5\serr{0.7}{0.5} & 3.2\serr{1.9}{0.8} & 2.3\serr{1.0}{0.6} & 3.5 & 48.5\\
140 & 1.4\serr{0.6}{0.4} & 3.2\serr{1.8}{1.0} & 2.3\serr{1.2}{0.7} & 3.3 & 65.8 \\

\end{tabular}
\end{ruledtabular}
\end{table*}

In summary, we have performed a direct search for the SM Higgs
boson decaying into $b$ jet pairs using data with integrated
luminosity of $0.97 \pm 0.06$~fb$^{-1}$ accumulated in Run II of
the CDF II detector. We observe no significant excess over the
background predicted by the SM and thus set 95\% C.L. upper limits
for the cross section of the Higgs boson production. The observed (expected)
combined limit of the weak boson associated Higgs boson production
cross sections set by this analysis for a Higgs boson mass of
115~GeV/$c^{2}$ is 20.4 (14.2) times the SM prediction. These results significantly
improve the combined limits on the allowed production
rate for the SM Higgs boson at the Tevatron.

We thank the Fermilab staff and the technical staffs of the participating institutions for their vital 
contributions. This work was supported by the U.S. Department of Energy and National Science 
Foundation; the Italian Istituto Nazionale di Fisica Nucleare; the Ministry of Education, Culture, Sports, 
Science and Technology of Japan; the Natural Sciences and Engineering Research Council of Canada; 
the National Science Council of the Republic of China; the Swiss National Science Foundation; the A.P. 
Sloan Foundation; the Bundesministerium f\"ur Bildung und Forschung, Germany; the Korean Science 
and Engineering Foundation and the Korean Research Foundation; the Science and Technology 
Facilities Council and the Royal Society, UK; the Institut National de Physique Nucleaire et Physique 
des Particules/CNRS; the Russian Foundation for Basic Research; the Comisi\'on Interministerial de
Ciencia y Tecnolog\'{\i}a, Spain; the European Community's Human Potential Programme; the Slovak
R\&D Agency; and the Academy of Finland.

\bibliography{apssamp}

\end{document}